       \documentclass[prb,aps,preprint]{revtex4-1}
       \usepackage{epsf,epsfig,textcomp}
       \def\br{{\bf r}}
       
       \def\bk{{\bf k}}
       \def\beq{\begin{equation}}
       \def\eeq{\end{equation}}
       
\begin{document}
       \title{Gapless metallic charge-density-wave phase \\
       driven by strong electron correlations}
       \author{Romuald Lema\'{n}ski}
       \address{Institute of Low Temperature and Structure Research, \\
       Polish Academy of Sciences, Wroc\l aw, Poland}
       \author{Klaus Ziegler}
       \address{Institut f\"ur Physik, Universit\"at Augsburg, Germany}
       \date{\today}
       
       \begin{abstract}
       We analyze the transformation from insulator to metal induced by thermal fluctuations 
       within the Falicov-Kimball model. Using the Dynamic Mean Field 
       Theory (DMFT) formalism on the Bethe lattice we find rigorously the temperature dependent Density of States ($DOS$) at half filling in the 
limit of high dimensions. At zero temperature ($T=0$) the system 
       is ordered to form the checkerboard pattern and the $DOS$ has the gap $\Delta$ at the Fermi level 
       $\varepsilon_F=0$, which is proportional to the interaction constant $U$. With an increase of $T$ the $DOS$ evolves in various ways that depend on $U$. 
For $U>U_{cr}$ the gap persists for any $T$  (then $\Delta >0$), so the system is always an insulator. However, if  $U < U_{cr}$, two additional subbands develop inside the gap. 
They become wider with increasing $T$ and at a certain $U$-dependent temperature $T_{MI}$ they join with each other at $\varepsilon_F$. Since above $T_{MI}$ the $DOS$ is positive at $\varepsilon_F$, 
we interpret $T_{MI}$ as the transformation temperature from insulator to metal. 
       It appears, that $T_{MI}$ approaches the order-disorder phase transition temperature $T_{O-DO}$ 
       when $U$ is close to $0$ or $ U_{cr}$, but $T_{MI}$ is substantially lower than $T_{O-DO}$ 
       for intermediate values of $U$. Having calculated the temperature dependent $DOS$ we study thermodynamic properties of the system starting from its free energy $F$. 
Then we find how the order parameter $d$ and the gap $\Delta $ change with $T$ and we construct 
the phase diagram in the variables $T$ and $U$, where we display regions of stability of four different phases: ordered insulator, ordered metal, disordered insulator and disordered metal.
Finally, we use a low temperature expansion to demonstrate the existence of a nonzero DOS 
at a characteristic value of $U$ on a general bipartite lattice.
       \end{abstract}
       \maketitle
       %\pacs{71.10.Fd, 71.30.+h, 71.45.Lr}

       \section{Introduction}
       One of the most successful methods for describing strongly correlated electron
       systems is the dynamical mean-filed theory (DMFT)
       \cite{GeorgesKotliarKrauthRozenberg,FreericksZlatic}. This formalism appears to be 
       particularly useful in studying the Falicov-Kimball model (FKM) \cite{FalicovKimball}, as it enables to get analytical, or high precision numerical results, which become exact in the limit of large dimensions.
       Most of the findings have been obtained in the high-temperature homogeneous phase
       \cite{GeorgesKotliarKrauthRozenberg}, but the ordered phase
       was also considered in a few papers     
       \cite{FreericksZlatic, BrandtMielsch,vanDongen,vanDongenLeinung,FreericksLemanski,GruberMacrisRoyerFreericks,ChenFreericksJones,HassanKrishnamurthy,MatveevShvaikaFreericks}.
       The results presented in these papers are remarkable, as they give a clear evidence that the static
       mean field theory is not an adequate tool for describing correlated electron systems. Indeed,
       physical quantities obtained using the static and dynamic mean field approach are substantially 
       different one from another. This discrepancy is particularly clearly demonstrated by Hassan and Krishnamurthy \cite{HassanKrishnamurthy}, and by Matveev, Shvaika and Freericks
       \cite{MatveevShvaikaFreericks}.
       Both teams analyzed the spinless FKM at half filling in the ordered charge-density-wave (CDW) phase having the form of checkerboard phase. Hassan and Krisnamurthy \cite{HassanKrishnamurthy} considered the square lattice and the Bethe lattice in the limit of infinite dimension and focused mostly on spectral properties, whereas Matveev, Shvaika and Freericks \cite{MatveevShvaikaFreericks} examined the hypercubic lattice in the limit of high dimensions and they focused mainly on transport properties.
       It is quite interesting that even though these studies were performed on different lattices, they lead to similar spectral properties of the model. Namely, in all the cases the energy spectrum has a gap at the Fermi level at $T=0$ and with an increase of $T$ two additional subbands develop inside the gap in such a way, that the density of states (DOS) at the Fermi level becomes positive still in the ordered phase (above a certain temperature $T_{MI}$), i.e. below the order-disorder transition temperature $T_{O-DO}$. 
In fact, the energy subbands developing inside the gap in the ordered checkerboard phase were
already noticed by Freericks and Zlati{\'c} \cite{FreericksZlatic}.
Here it is worthy to note that the Monte Carlo calculations performed on the 2D systems also give results similar to those obtained within DMFT \cite{MaskaCzajka,ZondaFarkasovsky}. 
       
       On the other hand, the data based on the static mean field theory calculations show that the gap disappears only at $T_{O-DO}$ \cite{FGebhard}. Indeed, according to a conventional mean field theory this gap gradually diminishes with an increase of temperature, but still persists until the CDW phase exists, i.e. until the order-disorder (O-DO) phase transition temperature $T_{O-DO}$ is reached \cite{FGebhard}.
       Surprisingly, the same conclusion was also formulated by van Dongen, who studied the FKM on the Bethe lattice using a different variant of DMFT \cite{vanDongen,vanDongenLeinung}. In fact, van Dongen derived analytical formulas on the temperature Green functions in the ordered phase, but he analyzed them only in the limiting cases of small and large coupling parameter $U$. Since in these two limits the gap is always present in the ordered phase, he concluded that it exists for any $U$. However this is in contradiction to the results reported in Refs. \cite{HassanKrishnamurthy,MatveevShvaikaFreericks}.
       
       Since the demonstration of the existence of the gapless ordered phase in 
       Refs. \cite{HassanKrishnamurthy,MatveevShvaikaFreericks} is quite surprising, but in the literature 
       still practically unnoticed result, in this contribution we develop studies of the subject. 
       Our purpose is to perform a more detailed analysis of spectral properties of the system focusing 
       mainly on intermediate values of the parameter $U$.  
       Following the approach derived by van Dongen \cite{vanDongen} we perform non-perturbative calculations 
       that allows us to reconstruct in a simple way the data obtained by Hassan and Krisnamurthy \cite{HassanKrishnamurthy} and to get analytical expressions for some characteristics of the spectrum not reported before. In addition, we calculated the electronic part of the specific heat and found that it behaves monotically around the temperature $T_{MI}$ of metal-insulator (MI) transformation. Hence, we conclude that the transformation is not a phase transition in the usual sense.
       
       Our analysis of the single electron energy spectrum of the spinless FKM 
       is based on exact formulas for the temperature-dependent $DOS$ $\rho(\varepsilon)$ 
       derived for the Bethe lattice within a version of the DMFT formalism derived by van Dongen \cite{vanDongen,vanDongenLeinung}. 
       There are two types of localized particles $A$ and $B$ in the system, whose densities $\rho_A$ 
       and $\rho_B$, respectively, are equal to each other and equal to $1/2$, 
       ($\rho_A=\rho_B=1/2$) and spinless electrons. 
       The localized particles may correspond, for example, to two different 
       components of an alloy. We focus on the half-filling case, when the density 
       of electrons $\rho_d=1/2$. Then the ground state has the checkerboard-type structure 
       composed of two interpenetrating sublattices $+$ and $-$, each of which is occupied 
       only by one type of particle: the sublattice $+$ by $A$ particles and 
       the sublattice $-$ by $B$ particles, respectively. Consequently, the density  
       $\rho^+_{A}$($\rho^-_{B}$) of particles $A(B)$ on the sublattice $+(-)$ 
       is equal to $1$ ($\rho^+_{A}=\rho^-_{B}=1$), whereas the density 
       $\rho^+_{B}$($\rho^-_{A}$) of particles $B(A)$ on the sublattice $+(-)$ 
       is equal to $0$ ($\rho^+_{B}=\rho^-_{A}=0$).
       
       With an increase of temperature the densities $\rho^+_{A}$, $\rho^-_{B}$ ($\rho^+_{A}=\rho^-_{B}$) 
       diminish below 1, while $\rho^+_{B}$, $\rho^-_{A}$ ($\rho^+_{B}=\rho^-_{A}$) increase above 0 and 
       in the disordered phase all these densities are equal to $1/2$. Then the quantity 
       $d =\rho^+_{A}-\rho^+_{B}=\rho^-_{B}-\rho^-_{A}$ is equal to 1 at $T=0$ 
       and equal to $0$ in the high-temperature, disordered phase,
       thus it is chosen to be the order parameter.
       It turns out that changes of $d$ cause significant changes in the $DOS$. 
       In particular, some energy states appear within the energy gap if $0<d<1$. 
       If it happens around the Fermi level, it corresponds to the MI transformation.
       
       In fact, the $DOS$ depends explicitly
       on the order parameter $d$ and its temperature dependence comes out
       entirely from the temperature dependence of $d$. Consequently, the order parameter $d(U;T)$ and 
       the $DOS$ $\rho(U,T; \varepsilon)$ are determined selfconsistently from the following 
       procedure. First we determine the $d$-dependent $DOS$ $\rho(U,d; \varepsilon)$ 
       and from that the free energy $F(U,d;T)$. 
       Next we find the temperature dependence of the order parameter $d(U;T)$ from minimization 
       of $F(U,d;T)$ over $d$. Then, we find the temperature dependent $DOS$ 
       $\rho(U,T; \varepsilon)$ by inserting $d(U;T)$ into $\rho(U,d; \varepsilon)$.
       And finally we calculate the internal energy $E(U,T)$,  
the energy gap $\Delta (U;T)$ and the value of $DOS$
       at the Fermi level $\rho(U,T; \varepsilon_F =0)$.
       
       The Hamiltonian we use is (see Ref. \cite{FreericksLemanski})
       \begin{eqnarray}
       \label{ham}
       H= & t\sum\limits_{<m,n>}d^+_m d_n+U\sum\limits_{m}w_mn^d_m &  \;
       \end{eqnarray}
       where $<m,n>$ means the nearest neighbor lattice sites $m$ and $n$, 
       $d_m$($d^+_m$) is an annihilation(creation) operator of itinerant electrons, 
       whereas $n^d_m$ is their particle number operator. 
       The quantity $w_m$ is equal to 1/2(-1/2) for the lattice site 
       occupied by the particle A(B), so the Coulomb-type on-site interaction between 
       itinerant electrons and the localized particles amounts $U/2(-U/2)$. 
       The hopping electron amplitude $t$ we henceforth set equal to one for our energy scale.
       
       We suppose that our results should be relevant to various experimental systems that display
       charge density or magnetic order such as for example $BaBiO_3$, $Ba_{1-x}K_xBiO_3$ 
       (see Ref. \cite{MatveevShvaikaFreericks} and the citations given therein) or 
       perovskite compounds $Ca(Mn_{3-x}Cu_x)Mn_4O_{12}$ and $TbBaCo_{2-x}Fe_xO_{5+\delta }$
       \cite{SzymczakSzewczyk,SzymczakSzymczak}.
       
       In the next section we provide a detailed analysis of the $DOS$ as a function of $d$ and $U$
       and in the section III we show the temperature dependence of the $DOS$. In the section III 
       we also discuss the relationship between the O-DO and MI transformations and present 
       the phase diagram of the system. 
Then the existence of a nonzero DOS at a characteristic value of $U$ is derived within a low temperature 
expansion on a general bipartite lattice (Sect. IV).   
       Finally, the last section contains some concluding remarks on our findings and a summary.

       \section{Density of states (DOS)}
       All physical properties analyzed in this paper are derived from 
       $\rho (U,d;\varepsilon )$ calculated from the Laplace transformation 
of the retarded Green function $G(U,d;\varepsilon )$ defined for complex $z$ 
with $Im(z)>0$ using the standard formula
       \begin{equation}
       \rho (U,d;\varepsilon)=-\frac{1}{\pi} ImG(U,d;\varepsilon +i0).
       \label{eq2}
       \end{equation} 
       In the remainder of this paper we will sometimes use simplified notations
       $G$ or $G(\epsilon)$ instead of $G(U,d; \epsilon)$ and $\rho(\epsilon)$
       instead of $\rho (U, d; \epsilon)$, respectively.
       
       For the two sublattice system one has
       \begin{equation}
       G(\varepsilon )=G^+(\varepsilon )+G^-(\varepsilon ), 
       \label{eq3}
       \end{equation}
       where the corresponding system of two equations for Green functions 
       $G^+(z)$ and $G^-(z)$ on the Bethe lattice reported by van Dongen
       \cite{vanDongen} is as follows.
       \begin{eqnarray}
       G^+(z)=\frac{z+\frac{1}{2}Ud -G^-(z)}{[z+\frac{1}{2}U -G^-(z)]
       [z-\frac{1}{2}U -G^-(z)]} \nonumber \\
       G^-(z)=\frac{z-\frac{1}{2}Ud -G^+(z)}{[z+\frac{1}{2}U -G^+(z)]
       [z-\frac{1}{2}U -G^+(z)]}, 
       \label{eq4}
       \end{eqnarray}
       
       At zero temperature $d=1$, so the system of eqs. (\ref{eq4}) 
       reduces to the following simple form
       \begin{eqnarray}
       G^+(z)=\frac{1}{z-\frac{1}{2}U -G^-(z)} \nonumber \\
       G^-(z)=\frac{1}{z+\frac{1}{2}U -G^+(z)}, 
       \label{eq5}
       \end{eqnarray}
       
       and the Green functions are expressed by the analytical formulas
       \begin{eqnarray}
       G^+(z)=\frac{4z^2-U^2-\sqrt{(4z^2-U^2)(4z^2-U^2-16)}}{4(2z-U)} \nonumber \\
       G^-(z)=\frac{4z^2-U^2-\sqrt{(4z^2-U^2)(4z^2-U^2-16)}}{4(2z+U)}. 
       \label{eq6}
       \end{eqnarray}
       
       It comes out from (\ref{eq6}) that the imaginary parts of $G^+(z)$ 
       and $G^-(z)$, so the $DOS$, have non-zero values within the intervals 
       $-(\sqrt{U^2+16})/2<\varepsilon <-U/2$ and $U/2<\varepsilon <(\sqrt{U^2+16})/2$. 
       Then the energy gap at the Fermi level is equal to $U$.
       Consequently, for any non-zero $U$ the system is an insulator at zero temperature.
       
       The situation is quite different at high temperatures, when the system is 
       in a disordered, homogeneous  state. In this case $d=0$,
       so $G^+(z)=G^-(z)=G(z)$ and the system of eqs. (\ref{eq4}) reduces to one
       polynomial equation of 3rd rank (eq. (\ref{eq7})) on $G(z)$. In fact, the equation (\ref{eq7}) 
       was first derived and analyzed already by Hubbard in his alloy analogy paper \cite{Hubbard}
(within the Hubbard-III-approximation of the Hubbard model). 
Then it was re-derived by Velicky et al. \cite{Velicky} and later on by van Dongen and Leinung \cite{vanDongenLeinung}.
       Here we rewrite it in the following form.
       \begin{equation}
       G^3-2zG^2+(1+z^2-U^2/4)G-z=0 
       \label{eq7}
       \end{equation}
       
       The equation (\ref{eq7}) has nontrivial analytic solutions that are significantly 
       different for small and large $U$. Consequently, for $U<U_{cr}=2$ there is no gap
       in the electronic energy spectrum, whereas for $U>U_{cr}=2$ there is the finite gap 
       at the Fermi level that increases with $U$. So the system is a conductor when $U$ 
       is smaller than the critical value $U_{cr} = 2$, otherwise it is an insulator.
       
\begin{figure}[h]
       \epsfxsize=13.5cm \epsfbox{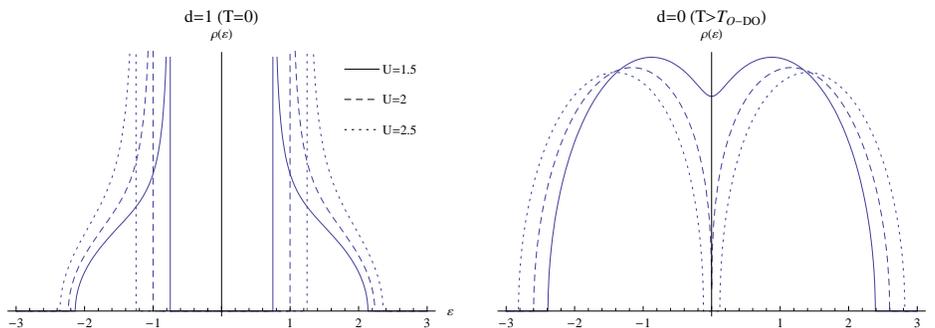}
       \caption{\it $DOS$ in the fully ordered phase ($d=1$) at $T=0$ (left panel) 
       and in the disordered phase ($d=0$) at $T>T_{O-DO}$ (right panel) for $U=1.5$ (the solid lines), 
       $U=2$ (the dashed lines) and $U=2.5$ (the dotted lines), respectively.}
       \label{figure1}
       \end{figure}

       In Fig. \ref{figure1} we display the $DOS$ in the ordered phase at $T=0$
       (left column) and in the disordered phase (right column) for a few representative
       values of $U$. It comes out that for $U>U_{cr}=2$
       the energy gap at the Fermi level persists in the disordered phase, then the system
       is an insulator. On the other hand, 
       for $U<U_{cr}$ the gap disappears in the high-temperature phase, so the order-disorder
       phase transition is accompanied by the insulator-metal transformation. However, it turns out
       that temperatures where these two transformations occur are usually different.
       
       The natural question that now arises is how the $DOS$ evolves with temperature starting 
       from $T=0$ and ending at high temperature, where the system is in the disordered phase. 
As we mentioned in the Introduction, preliminary studies of the $DOS$ for the ordered phase 
at finite temperatures were already reported in the review paper by Freericks and Zlatic \cite{FreericksZlatic}.
Then this problem was examined by Hassan and Krishnamurthy \cite{HassanKrishnamurthy}
and independently by Matveew, Shvaika and Freericks \cite{MatveevShvaikaFreericks}. 
In all these papers the authors calculated $\rho (\varepsilon )$ using the method of summation 
over Matsubara frequencies.
Here we get similar results using a different method. Namely, we solve the system 
of eqs. (\ref{eq4}) for arbitrary $d$ 
       and then we calculate $\rho (\varepsilon )$ from eqs. (\ref{eq2}) and (\ref{eq3}). 
       In fact, the system of eqs. (\ref{eq4}) reduces to the polynomial equation of 5rd rank 
       on $G^+(z)$ (see eq. (\ref{eq8})) or $G^-(z)$ (not displayed, but knowing $G^+(z)$ 
one can find $G^-(z)$ from (\ref{eq4})). 
\begin{equation}
a_0+a_1G^++a_2(G^+)^2+a_3(G^+)^3+a_4(G^+)^4+a_5(G^+)^5=0
\label{eq8}
\end{equation}
The coefficients $a_0$, $a_1$, $a_2$, $a_3$, $a_4$, $a_5$ are functions of $z$, $U$ and $d$.
Since the expressions on these coefficients are rather lengthy, we put them into Appendix \ref{sect:a}.

       \begin{figure}[h]
       \epsfxsize=8cm \epsfbox{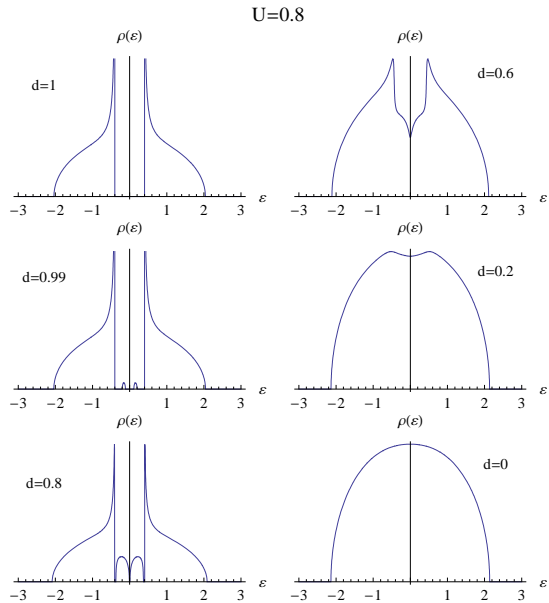}
       \caption{\it Evolution of the $DOS$ with a change of the order parameter $d$ 
       from the fully ordered phase at $T=0$ ($d=1$) to the high-temperature, disordered phase ($d=0$) 
       for $U=0.8$. In this case the insulator-metal transformation occurs in the system 
       (at $d=4\sqrt{21}/25\approx 0.733$).}
       \label{figure2}
       \end{figure}
       \begin{figure}[h]
       \epsfxsize=8cm \epsfbox{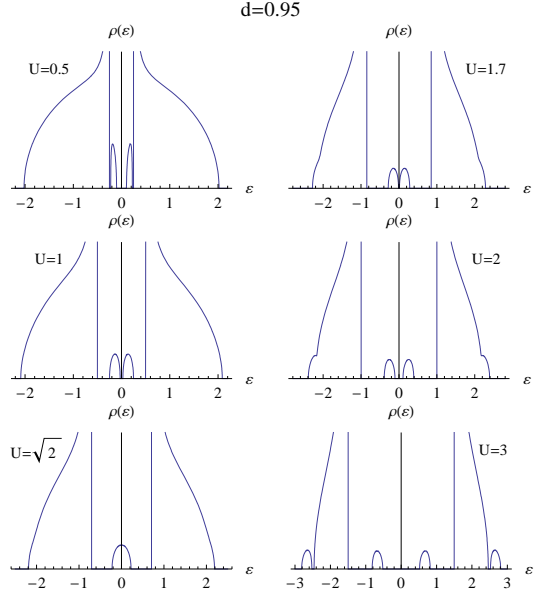}
       \caption{\it Evolution of the $DOS$ with $U$ for the fixed value of the order parameter $d=0.95$. }
       \label{figure3}
       \end{figure}
       
The resulting $DOS$ is displayed in Fig. \ref{figure2} for $U=0.8$ and a set of $d$ values, whereas 
       in Fig. \ref{figure3} for $d=0.95$ and a set of $U$ values. 
       By viewing Fig. \ref{figure2} one can see how $\rho (\varepsilon )$ 
       evolves when the system undergoes the MI transformation and by viewing Fig. \ref{figure3}
       one can notice how the process of filling in the gap starts up when the order parameter
       $d$ begins to be less than one. As we already mentioned before this filling is quite surprising,
       as being completely different from the expectations based on the conventional mean field theory \cite{FGebhard}.  
       Indeed, according to this theory the process of closing the gap is due to gradual increase 
       in the width of two DOS subbands: one lying just below and the other just above the Fermi level. 
       As a result, the upper edge of the valency band and the lower edge of the conduction band converge 
       to each other if, and only if $d=0$.
       
       On the other hand, our results confirm findings reported 
in Refs. \cite{HassanKrishnamurthy,MatveevShvaikaFreericks} that the filling of the gap occurs 
       due to two additional subbands developing inside the gap. 
       These additional subbands are located symmetrically with respect to the Fermi level
       and their initial positions depend on $U$ (for $d$ just below 1). With a decrease of $d$ 
       the width of the subbands increases, and they merge together to form one band 
       at certain $d_{crit}(U)$, if $0\leq U<U_{cr}=2$.
       
       From our calculations it was easy to obtain a simple analytical formula for DOS 
       at the Fermi level $\rho(\varepsilon_F=0)$ as a function of $U$ and $d$.
       Indeed, it appears that at this special point the polynomial in eq. (\ref{eq8})
factorizes, so that the eq. (\ref{eq8}) has the following simple form.
\begin{equation}
U (4 G^+ - 4 (G^+)^3 + 2 d U + G^+ U^2) (8 d G^+ + 4 U + 4 (G^+)^2 U - U^3)=0
\label{eq8a}
\end{equation}
Then the eq.  (\ref{eq8}) can be solved analytically and the resulting DOS is as follows.
       \begin{equation}
       \rho(\varepsilon_F)\equiv\rho(U,T;\varepsilon_F=0)=\frac{1}{\pi}Im(\frac{\sqrt{4d^2-4U^2+U^4}}{2U})
       \label{eq9}
       \end{equation}
       Hence it follows that inside the whole interval $0<U<2$ the system is metallic 
       (i.e. $\rho(\varepsilon_F)>0$) not only in the disordered phase where $d=0$
       (then $\rho(\varepsilon_F)=\frac{\sqrt{4-U^2}}{2\pi}$), but also
       in the ordered phase, if only $d<d_{crit}(U)=\frac{U}{2}\sqrt{4-U^2}$.

       Moreover, at $U=\sqrt{2}$ the maximum value $d_{crit}(U)=1$ is attained,
       so the system is then metallic even  for $d$ infinitesimally close to the limit $d=1$,
       that corresponds to the fully ordered phase at $T=0$.
       
       Having the formula for DOS derived from the system of eqs. \ref{eq4} we are also able 
       to analyze the insulating phase characterized by its energy gap at the Fermi level 
       $\Delta(\varepsilon_F)$ (then obviously $\rho(\varepsilon_F)=0$). If $U\geq 2$ the system 
       is an insulator both in the disordered and ordered phase for any $d$. On the other hand,
       if $0<U<2$, then it is an insulator only for $d_{crit}(U)\leq d\leq 1$.
       As we already mentioned before, at $T=0$, i.e. in the fully ordered phase ($d=1$) 
       one has $\Delta(\varepsilon_F)(U)=U$.
       However, it appears that $\Delta(\varepsilon_F)$ is not a continuous function of $d$
       at $d=1$. Indeed, when $d<1$ and $d\rightarrow 1$ (i.e. $T=0^+$) we got the following 
       analytical formula 
       \begin{equation}
       \Delta(\varepsilon_F)=|\frac{\sqrt{1+4U^2}-U^2-1}{U}|, \; d\rightarrow 1 \; (d<1)
       \label{eq10}
       \end{equation}
       and also the analytical expression for $d=0$ 
       ($T>T_{O-DO}$) (see Eq. (\ref{eq11})).
       
       \begin{eqnarray}
        & 0 & \; d=0, \; U<2 \nonumber \\
       \Delta(\varepsilon_F)=  &  & \nonumber \\
       & \; \sqrt{10+U^2-2\frac{1+\sqrt{(1+2U^2)^3}}{U^2}}, & \; d=0, \; U>2 
       \label{eq11}
       \end{eqnarray}
       As far as we know, the formulas given in (\ref{eq10}) and (\ref{eq11}) 
       were not published before.
       
\begin{figure}[h]
       \epsfxsize=8cm \epsfbox{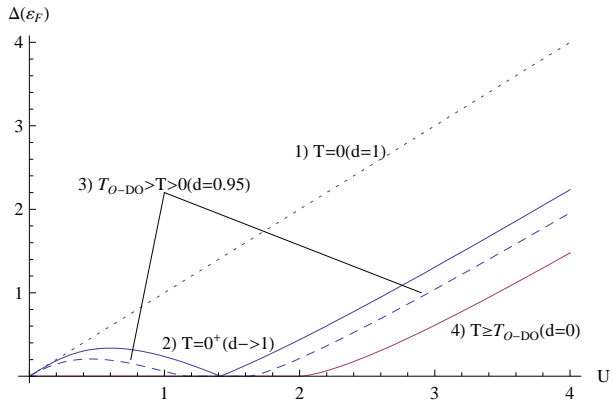}
       \caption{\it Energy gap $\Delta (\varepsilon_F)$ at the Fermi level as a function of U for a few fixed values of the order parameter d.}
       \label{figure4}
       \end{figure}

       In Fig. \ref{figure4} we display how the energy gaps $\Delta(\varepsilon_F)$
       change with $U$ for a set of few fixed values of $d$.
       At $T=0$ $(d=1)$ $\Delta(\varepsilon_F)(U)$ is represented by the straight dotted line
$\Delta(\varepsilon_F)(U)=U$.
       However, for $d<1$, but $d$ being infinitesimally close to 1 the function 
       $\Delta(\varepsilon_F)(U)$ behaves non-monotonically. Starting from zero at $U=0$ 
       it first increases, attains its local maximum equal to $\frac{4\sqrt{18-2\sqrt{17}}+\sqrt{17}-15}{2\sqrt{14-2\sqrt{17}}}\approx 0.33675$
       at $U=\frac{\sqrt{\frac{7-\sqrt{17}}{2}}}{2}\approx 0.6$
       and then goes down to 0 at $U=\sqrt{2}$.
       In the opposite limit of the homogeneous phase ($d=0$) one has $\Delta(\varepsilon_F)(U)=0$
       for $U\leq 2$ and the curve $\Delta(\varepsilon_F)(U)$ starts to rise up for $U\geq 2$
       according to the formula (\ref{eq11}).
       The behavior of $\Delta(\varepsilon_F)(U)$ between these two limits is represented
       in Fig. \ref{figure4} for $d=0.95$ by the dashed line.

Note also that when $U\rightarrow 0$, then from the formula (\ref{eq10}) one has 
$\Delta (\varepsilon_F)\rightarrow |U|$, and when $U\rightarrow \infty $, then from (\ref{eq11}) 
one gets $\Delta (\varepsilon_F)\rightarrow |U|-2$. This is why the exact analytical calculations performed in the limiting cases of small and 
large $U$ by van Dongen \cite{vanDongen} could not detect the gapless checkerboard phase.
       
       \section{Order-disorder versus insulator-metal transition}
\begin{figure}[h]
       \epsfxsize=8cm \epsfbox{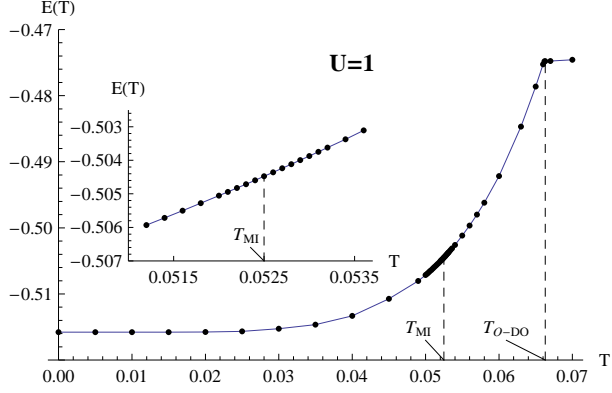}
       \caption{\it Internal energy $E (U,T)$ as a function of $T$ for $U=1$. The temperature interval
close to $T_{MI}$ is displayed in the inset. Drawn lines are guides to the eye.}
       \label{figure5}
       \end{figure}

       Having calculated $\rho(U,d;\varepsilon)$ we can determine the free energy 
       functional using the formula \cite{FreericksGruberMacris,ShvaikaFreericks}
       \begin{eqnarray}
       F(U,d,T)=T\int_{-\infty}^{\infty}{d\varepsilon \rho(U,d;\varepsilon)
       ln\frac{1}{1+exp(-\varepsilon/k_BT)}}+T(\frac{1+d}{2}ln\frac{1+d}{2}+\frac{1-d}{2}ln\frac{1-d}{2})
\label{eq12}
       \end{eqnarray}
       and by minimizing $F(U,d,T)$ over $d$ we can find the order parameter $d(U;T)$.
       Then, by inserting $d(U;T)$ into $\rho(U,d;\varepsilon)$ we get $\rho(U,T;\varepsilon)$. 
       Next, from $\rho(U,T;\varepsilon)$ we determine the internal energy $E(U,T)$ 
using the standard formula (\ref{eq13})
\begin{equation}
       E(U,T)=\int_{-\infty}^{\infty}{d\varepsilon \rho(U,T;\varepsilon)
       \frac{\varepsilon }{1+exp(\varepsilon/k_BT)}}
\label{eq13}
       \end{equation}
and the temperature dependence of two quantities characterizing the MI transformation: the energy gap 
       $\Delta (U;T)$ and the  $DOS$ at the Fermi level $\rho(\varepsilon_F=0;T)$.
We display $E(U=1,T)$ as a function of $T$ in Fig. \ref{figure5}, where it can be seen
that this function has a kink at $T_{O-DO}$, but no kink or any noticable anomaly at $T_{MI}$.
This is why we conclude that the metal-insulator transformation at $T_{MI}$
is not a phase transiton in the usual sense. On the other hand, at $T_{O-DO}$ the system undergoes 
a typical order-disorder phase transition.

\begin{figure}[h]
       \epsfxsize=8.5cm \epsfbox{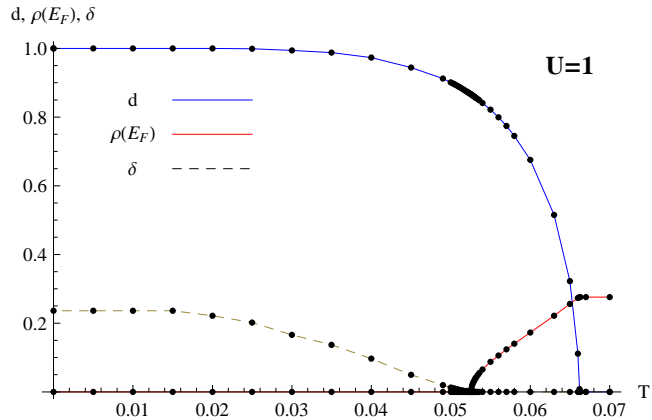}
       \caption{\it Temperature dependence of the order parameter $d$, $DOS$ $\rho(\varepsilon_F)$ 
       at the Fermi level and the relative value of energy gap $\delta$ for $U=1$. 
       Drawn lines are guides to the eye.}
       \label{figure6}
       \end{figure}

       The temperature dependencies of $d$, $\rho(\varepsilon_F)$ and $\delta$ for $U=1$ 
are displayed in Fig \ref{figure6}, where $\delta=\Delta(T) /\Delta (T=0)$ is the relative value 
of the gap. Note, that $\delta$ has a jump at $T=0$ 
because $\Delta (T=0)=U$ but, as it comes from Eq. (\ref{eq10}), $\Delta (T=0^+)<U$. 
Then, for $U=1$ one has $\delta (T=0)=1$ and $\delta (T=0^+)=\sqrt 5-2\approx 0.236$.
Obviously, the energy gap $\Delta$, so do $\delta$ is positive in the insulating phase, 
i.e. for $T<T_{MI}$ and is equal to zero in the metallic phase. 
       On the other hand, $\rho(\varepsilon_F)$ is equal to zero in the insulating phase, 
       but is positive in the metallic phase.
       
       By viewing Fig. \ref{figure6} one can see that $T_{MI}\approx 0.052$ and $T_{O-DO}\approx 0.0662$
       for $U=1$, so $T_{MI}$ is substantially smaller than $T_{O-DO}$. One can also notice that
       MI transformation occurs when the order parameter $d\approx 0.9$, so $d$ is still
       close to its maximum value 1. Another interesting observation is that $\rho(\varepsilon_F)$   
       clearly increases with temperature up to the maximum value $\rho(T=T_{O-DO};\varepsilon_F)$ 
       attained at $T_{O-DO}$ and this value is preserved for higher temperatures.
       
\begin{figure}[h]
       \epsfxsize=9cm \epsfbox{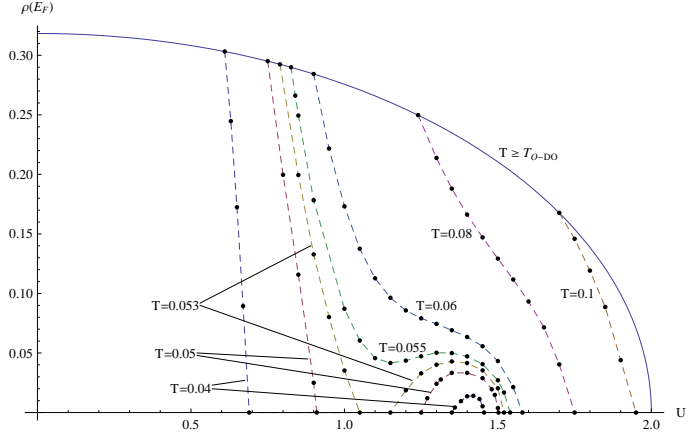}
       \caption{\it $DOS$ $\rho(\varepsilon_F)$ as a function of $U$
       for a representative set of temperatures. Drawn lines are guides to the eye.}
       \label{figure7}
       
       \end{figure}

\begin{figure}[h]
       \epsfxsize=8.5cm \epsfbox{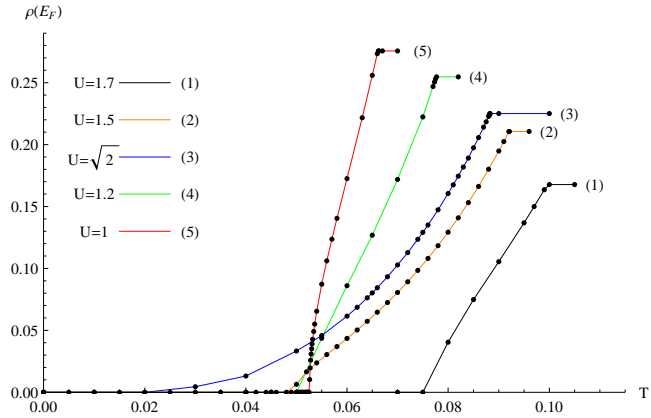}
       \caption{\it $DOS$ $\rho(\varepsilon_F)$ as a function of $T$
       for a few $U$ values. Drawn lines are guides to the eye.}
       \label{figure8}
       \end{figure}

       After inserting $d(U;T)$ into (\ref{eq9}) we get $\rho(\varepsilon_F)$
       as a function of $U$ and $T$. This function is quite non-trivial as it can be seen
       in Figs. \ref{figure7} and \ref{figure8}. In Fig. \ref{figure7} we display
       $\rho(\varepsilon_F)$ as a function of $U$ for a set of fixed temperatures
       and in Fig. \ref{figure7} one can observe $\rho(\varepsilon_F)$ as a function of $T$
       for a few $U$ values.
       

\begin{figure}[h]
       \epsfxsize=12cm \epsfbox{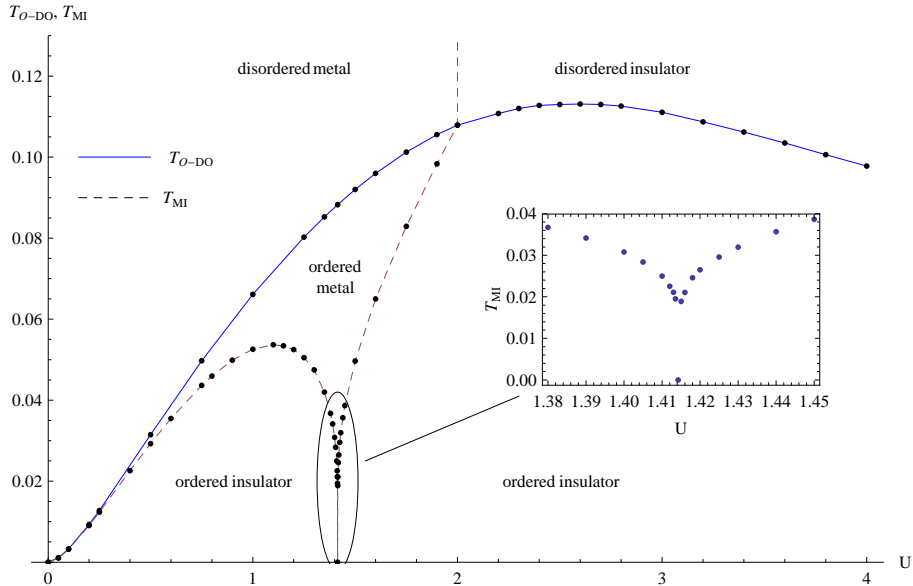}
       \caption{\it Phase diagram of the system at finite temperatures.
       $T_{O-DO}$ is the order-disorder transition temperature 
       (displayed by points and the solid line, that is a guide to the eye; online in blue) and $T_{MI}$ 
       is the metal-insulator transformation (displayed by points and the dashed line that is a guide 
       to the eye). The solid line separates the disordered phase (the upper part of
       the diagram) from the ordered phase (the lower part of the diagram) and the dashed line separates 
       the insulating phase (on the right) from the conducting phase (on the left).
       In the middle of the diagram there is an area of stability of the ordered metallic phase}
       \label{figure9}
       \end{figure}

       After collecting the data on $T_{O-DO}$ and $T_{MI}$ for a representative 
       set of $U$ values we constructed the phase diagram of the system that is displayed 
       in Fig. \ref{figure9}. Let us note that in this diagram the region of ordered insulator 
       phase located below $T_{O-DO}$ (continuous) line consists of two parts corresponding to insulating 
       phases separated by an ordered metallic phase. This is quite unexpected finding obtained neither within the conventional 
       mean field theory \cite{FGebhard}, 
       nor through the exact procedure of expanding in series for large or small $U$ values \cite{vanDongen}. In fact, 
       the finding is not inconsistent with the result obtained 
       by Van Dongen \cite{vanDongen}, as indeed, for small and large $U$ the gap exists in the ordered phase up to $T_{O-DO}$. 
       However, for intermediate $U$ 
       values the metallic ordered phase appears    below $T_{O-DO}$ down to $T_{MI}$. What's more, for $U=\sqrt{2}$ the metallic 
       phase can be stable down to $T_{MI}=0$.
       
       The phase diagram displayed in Fig. \ref{figure9} is almost identical to the one 
       presented in Ref. \cite{HassanKrishnamurthy}.
       However, there is a substantial difference between the two diagrams at $U=\sqrt{2}$,
       where in our case the end point of the homogeneous phase is at $T=0$,
       whereas in Ref. \cite{HassanKrishnamurthy}
       it lies slightly above $T=0$. This is due to difference in calculation techniques used
       in the two cases. We were able to fix this end point at $T=0$ using the analytical
       formula (\ref{eq9}). Then the question arises about quantum effects related to the MI
       transformation for this particular value of $U$. In order to clarify this point 
       some additional studies need to be done.

\section{Low-temperature expansion on the lattice}       
       
Now we consider a grand-canonical ensemble with the spinless Falicov-Kimball  
Hamiltonian (\ref{ham}) on a $d$-dimensional bipartite lattice. To distinguish
the following calculation from the previous one on the Bethe lattice, we 
introduce lattice coordinates $r$, $r'$. With $n_r=w_r+1/2$, where the absence 
(presence) of a heavy fermions at site $r$ is represented as classical binary number 
$n_r=0$ ($n_r=1$), we can write for the Hamiltonian matrix
\beq
{\tilde H}_{r,r'}=h_{r,r'}-(\mu - U n_r)\delta_{r,r'}
\eeq
with the chemical potential $\mu$. At half filling we have for the latter $\mu=U/2$.

According to Ref.~[\onlinecite{ates05}], the heavy particles are distributed by the thermal distribution at the inverse temperature $\beta=1/k_B T$
\begin{equation}
P(\{n_\br\})=e^{\beta\mu\sum_r n_r}\det\left({\bf 1}+e^{-\beta{\tilde H}}\right)/Z , \ \ \ 
Z=\sum_{\{ n_r=0,1\}}e^{\beta\mu\sum_r n_r}\det\left({\bf 1}+e^{-\beta{\tilde H}}\right)
\end{equation}
which we can approximate by an Ising distribution as %\cite{ates05}:
\begin{equation}
e^{\beta\mu\sum_r n_r}\det\left({\bf 1}+e^{-\beta{\tilde H}}\right)
=e^{-2\beta t^2\sum_{<r,r'>}(n_r-1/2)(n_{r'}-1/2)+o(t^3)}
\ .
\end{equation}
Adding or removing a heavy particle from the ground state (staggered configuration) appears then 
with the weight $w^{-\beta}$, where
\begin{equation}
w\sim e^{-2 t^2/U}
\ .
\end{equation}
This provides us a low temperature expansion for the density of states (cf. App. \ref{sect:b})
by adding or removing particles from the groundstate configuration:
\beq
\rho =w^{-\beta}(U^3g^2/4)\delta(1-gU^2/2)+o(w^{-2\beta})
\ .
\label{dos_contr}
\eeq
Thus, in order $w^{-\beta}$ we have a Dirac delta function for the DOS
which is peaked at $U^2=2/g$ and has a weight $w^{-\beta}U^3 g^2/4$, where the parameter
can be calculated as an integrals for a given lattice with known hopping term $h_k$:
\beq
g=\int_\bk \frac{1}{U^2/4 %+\delta^2
+|h_k|^2}
\ .
\eeq
The contribution to the DOS in Eq. (\ref{dos_contr}) vanishes with decreasing temperature, similar to the
DOS in Figs. \ref{figure7}, \ref{figure8}. With increasing temperature we must include higher order terms in $w^{-\beta}$
which might lead to a broadening of the DOS around $U=\sqrt{2/g}$. These results indicate that the
singular DOS around a special value of $U$ in Fig. \ref{figure7},  is not an artifact of the DMFT
or the Bethe lattice but a general feature of the FK model on any bipartite lattice.

       \section{Final remarks and conclusions}
       Here we focus on the quantitative analysis of a relationship between the degree of disorder 
       in a correlated electron system and the transformation from insulator to metal. 
       Using exact formulas for the temperature-dependent $DOS$ for the FK model on the Bethe lattice 
       we demonstrate the effect of closing of the energy gap in the $DOS$ in the insulating phase 
       (for not too large $U$) and then of increasing of the $DOS$ value at the Fermi level 
       in the metallic phase with an increase of degree of disorder. 
       Our results confirm and extend the findings presented 
        in Refs. \cite{HassanKrishnamurthy,MatveevShvaikaFreericks}.
       
       One of the most surprising conclusions drawn from these studies is that an increase of disorder 
       may lead to a closure of the energy gap still before the system transforms into a completely 
       disordered phase.
       In view of this result, we suggest a re-examination of those experiments, in which  
       transition temperatures $T_{O-DO}$ and $T_{MI}$ are found to be the same \cite{RajanAvignonFalicov,SzymczakSzewczyk,SzymczakSzymczak}.
       But one should keep in mind that the distinction between $T_{O-DO}$ and $T_{MI}$ 
       can be difficult to detect in some systems, as a clear difference between these temperatures
       was found only in a relatively narrow range of values of the parameter $U$.
       An additional difficulty is that just above $T_{MI}$ the $DOS$ at the Fermi level is still small, 
       as only above $T_{MI}$ it begins to rise with temperature, starting from zero and reaching 
       a maximum value at $T_{O-DO}$ (see Fig. \ref{figure7}).
       Therefore, we expect, that one will be able to notice a difference between 
       $T_ {MI}$ and $T_ {O-DO}$ only in precise enough experiments.
       
       As we have demonstrated within a low temperature expansion, the results emerging from the DMFT calculation on the Bethe lattice might be quite general.
       The reason is, that the FK model, called by some authors the simplified Hubbard model
       \cite{MichielsenRaedt}, contains basic ingredients that are present in many other models 
       of correlated electron system. 
       On the other hand, properties of the $DOS$ relevant for these studies, such as
       existence of the gap in the homogeneous phase for sufficiently large $U$ and closing the gap
       with decreasing $U$ in the homogeneous phase but not in the ordered phase, are common 
       for all examined lattices (hypercubic 1D, 2D, 3D and the infinite D, as well as the Bethe lattice
       in the infinite D limit) 
       \cite{vanDongen,vanDongenLeinung,MichielsenRaedt,MaskaCzajka,ZondaFarkasovsky}.
       
       Interestingly enough, there are some similarities between our phase diagram displayed 
       in Fig. \ref{figure6} and the phase diagram found for the Hubbard model with 
       disorder \cite{ByczukVollhardt}. 
       In fact, we cannot directly compare our results with those reported in Ref. \cite{ByczukVollhardt}, 
       as these latter were obtained not for the FK model but for the Hubbard model, and only 
       at zero temperature. However, in these two cases the same sort of phases appear on the phase diagram, 
       only insulating phases survive for large $U$ and the ordered metallic phase occupies a relatively small region in the phase diagram.
       
       Finally, let us hope that the existence of gapless checkerboard-type charge density wave phase
found first for the FK model will be confirmed by studies for on the Hubbard model and other models of strongly correlated electrons.
       
       \acknowledgments
       We express our best thanks to K. Byczuk and J.K. Freericks for useful discussions on some issues raised in this paper and for critical reading of the manuscript.

\appendix       
\section{Coefficients of the polynomial given in Eq. (\ref{eq8})}
\label{sect:a}
Here are the coefficients $a_0$, $a_1$, $a_2$, $a_3$, $a_4$, $a_5$ given in eq. (\ref{eq8}) 
that are obtained from the transformation of the system of eqs. (\ref{eq4}).
\begin{eqnarray}
&a_0=-2 (4 z^2 - U^2) (8 z^3 + 4 d z^2 U - d U (-4 + U^2) - 2 z (4 + U^2))& \nonumber \\ 
&a_1=64 z^6 + 192 d z^3 U - 48 d z U^3 - 16 z^4 (-8 + 3 U^2) + 
U^2 (16 + 16 d^2 - U^4) + 4 z^2 (-32 - 8 U^2 + 3 U^4)& \nonumber \\
&a_2=-16 (16 z^5 + 20 d z^2 U - 8 z^3 U^2 - d U (2 + U^2) + z (-4 + U^4))& \nonumber \\
&a_3=8 (48 z^4 + 24 d z U + U^4 - 16 z^2 (1 + U^2))& \nonumber \\
&a_4=-32 (8 z^3 + d U - 2 z (1 + U^2))\nonumber \\
&a_5=64 z^2 - 16 U^2 \nonumber
\end{eqnarray}

\section{Green's function on the bipartite lattice}
\label{sect:b}

Then the Green's function of the light fermions reads as an average with
respect to a grand-canonical distribution of the heavy fermions
\beq
G=\langle (H-i\delta)^{-1}\rangle
\equiv \sum_{\{n_\br=0,1\}}P(\{n_\br\})(H-i\delta)^{-1}
\ .
\label{gf00}
\eeq
At half-filling, where $\mu=U/2$, the ground state of the heavy
particles on a bipartite lattice is a staggered (or generalized checkerboard)
configuration. Using a sublattice representation for the hopping of the light 
fermions, we obtain
\beq
{\bar H}=\pmatrix{
U/2 & h \cr
h^T & -U/2 \cr
}
\ ,
\eeq
where the sublattice 1 (2) has the effective potential $U/2$ ($-U/2$). Here we 
have assumed that the hopping is only between nearest neighbors. Therefore, 
the hopping terms are $h$, $h^T$ in the off-diagonal elements of our sublattice 
matrix. Now we apply a Fourier transformation on the translational invariant 
sublattice to get as Fourier components $2\times 2$ matrices
\beq
{\bar H}_k=\pmatrix{
U/2 & h_k \cr
h_k^* & -U/2 \cr
}
\eeq
with the two-band dispersion $E_k=\pm\sqrt{U^2/4+|h_k|^2}$. The sum over other configurations in 
(\ref{gf00}) is now an expansion in powers of a weight $w^{-\beta}$.
This implies for the Green's function
\beq
G=\sum_{\{n_\br=0,1\}}P(\{n_\br\})({\bar H}+Un-i\delta)^{-1}
\eeq
\beq
=({\bar H}-i\delta)^{-1}+w^{-\beta}
\sum_r\left[({\bar H}+V_r-i\delta)^{-1} +({\bar H}+W_r-i\delta)^{-1}\right]
+o(w^{-2\beta})
\eeq
with
\beq
V_r=\pmatrix{
-U & 0 \cr
0 & 0 \cr
}, \ \ \ 
W_r=\pmatrix{
0 & 0 \cr
0 & U \cr
}
\ .
\eeq
The latter expressions mean that $V_r$ ($W_r$) removes (adds) a heavy particle at site $r$ on sublattice
1 (2).
The expressions $({\bar H}+V_{r}-i\delta)^{-1}$, $({\bar H}+W_{r}-i\delta)^{-1}$ can be easily
computed by using the identity %\cite{ziegler85}
\beq
(A+\eta)^{-1}=A^{-1} -A^{-1}
({\bf 1}+\eta A^{-1})^{-1}_S\eta A^{-1}
\ ,
\eeq
where $S$ refers to the projection of the matrix space with nonzero $\eta$. In our case
$S$ is just the single site $r$, such that this identity reads with ${\bar G}=({\bar H}-i\delta)^{-1}$
\beq
({\bar H}+V_r-i\delta)^{-1}_{r'j,r'j}={\bar G}_{r'j,r'j}
-{\bar G}_{r'j,r1}\frac{-U}{1-U{\bar G}_{r1,r1}}{\bar G}_{r1,r'j}
\eeq
\beq 
({\bar H}+W_r-i\delta)^{-1}_{r'j,r'j}={\bar G}_{r'j,r'j}
-{\bar G}_{r'j,r2}\frac{U}{1+U{\bar G}_{r2,r2}}{\bar G}_{r2,r'j}
\eeq
The elements of the Green's function ${\bar G}$ can be evaluated from their Fourier components as
\beq
{\bar G}_{r1,r1}=(U/2+i\delta)g , \ \ \ 
{\bar G}_{r2,r2}=(-U/2+i\delta)g, \ \ \ g=\int_\bk \frac{1}{U^2/4+\delta^2+|h_k|^2}
\label{int0}
\eeq
such that
\beq
({\bar H}+V_r-i\delta)^{-1}_{r'j,r'j}={\bar G}_{r'j,r'j}
-{\bar G}_{r'j,r1}\frac{-U}{1-gU^2/2-iUg\delta}{\bar G}_{r1,r'j}
\eeq
\beq
={\bar G}_{r'j,r'j}
+\frac{U(1-gU^2/2+iUg\delta)}{(1-gU^2/2)^2+U^2g^2\delta^2}{\bar G}_{r'j,r1}{\bar G}_{r1,r'j}
\eeq
\beq 
({\bar H}+W_r-i\delta)^{-1}_{r'j,r'j}={\bar G}_{r'j,r'j}
-{\bar G}_{r'j,r2}\frac{U}{1-gU^2/2+iUg\delta}{\bar G}_{r2,r'j}
\eeq
\beq
={\bar G}_{r'j,r'j}
-\frac{U(1-gU^2/2-iUg\delta)}{(1-gU^2/2)^2+U^2g^2\delta^2}{\bar G}_{r'j,r2}{\bar G}_{r2,r'j}
\ .
\eeq
Since we have a gap $U$, the Green's function ${\bar G}$ is real in the limit $\delta\to0$.
Therefore, the density of states reduces to
\beq
\rho=\frac{1}{\pi}\lim_{\delta\to0}{\rm Im}G_{rj,rj}
=w^{-\beta}\frac{1}{\pi}\lim_{\delta\to0}\frac{U^2g\delta}{(1-gU^2/2)^2+U^2g^2\delta^2}{\bar G}^2_{r'j,r'j}
+o(w^{-2\beta})
\eeq
\beq
=w^{-\beta}U{\bar G}^2_{r'j,r'j}\delta(1-gU^2/2)+o(w^{-2\beta})
\eeq
with the Dirac delta function $\delta(x)$. 

\end{document}